\documentclass[12pt]{article}
\usepackage{amsfonts}
\usepackage{a4wide}

\newcommand{\be}{\begin{equation}}
\newcommand{\ee}{\end{equation}}

\newcommand{\cZ}{{\cal Z}}
\newcommand{\g}{\mathfrak{g}}
\newcommand{\ft}{\mathfrak{t}}

\newcommand{\cC}{{\cal C}}

\newcommand{\CP}{\mathbb{C}\mathbb{P}}

\newcommand{\bV}{\mathbb{V}}

\newcommand{\C}{\mathbb{ C}}

\newcommand{\rd}{{\rm d}}

\newcommand{\cO}{{\cal O}}

\newcounter{note}

\begin{document}

\title{Two twistor descriptions of the isomonodromy problem}
\author{N.M.J Woodhouse}
\maketitle
\begin{abstract}
The connections between Hitchin and Mason's 
twistor descriptions of the isomonodromy problem are explored.
\end{abstract}

\section{Introduction}
Twistor theory was first explored by Penrose in his investigation of
the role of holomorphicity and conformal symmetry in relativistic
quantum field theory. The primary aim, to find a new route to the
quantization of gravity, has yet to be fully achieved; but through the
work of Ward, Hitchin, Mason, and others, the underlying geometry has
provided a unifying framework for the study of integrable
systems (see \cite{MW} for a review).  
It sheds light on the connections between
\begin{itemize}
\item integrable systems of partial differential equations;

\item real and complex geometries with symmetry; and

\item isomonodromic families of ordinary differential equations.
\end{itemize}
In this note, I shall concentrate on the third of these, and will
explain the connections between two different twistor representations
of isomonodromy, due, respectively, to Hitchin and Mason. The former
construction links isomonodromy to problems in differential
geometry; the latter gives a tool for the the systematic study of the
way in which the isomonodromic deformation equations arise from the
dimensional reduction of integrable systems.  The route from the first
construction to the second has not appeared elsewhere.

\section{Conformal reductions of the self-dual Yang-Mills equations}

Ward showed \cite{W} that there is a correspondence between, on the one hand,
anti-self-dual solutions to the Yang-Mills equations on a suitable
region of complex space-time and, on the other, holomorphic vector
bundles over a corresponding subset $U\subset \CP_3$ (complex
projective 3-space).  The Yang-Mills field is a connection $D$ on a
trivial bundle and the anti-self-duality condition is that its curvature
should vanish on a special three-dimensional family of null 2-planes
($\alpha$-planes). The set of $\alpha$-planes 
is the {\em twistor space} of complex
space-time---it is identified with a subset of $\CP_3$, in a way that
is natural in the sense that the action of the proper conformal group
in space-time corresponds to the action of the isomorphic group ${\rm
PGL}(4,\C)$ on $\CP_3$.  Ward's construction maps $D$ to the vector bundle
$E\to U$ whose fibres are the spaces of solutions to the linear
equations $Ds=0$ over the $\alpha$-planes.  The remarkable and
non-trivial fact is that the construction is reversible: $D$ can be
uniquely recovered from $E$, with no other data required---a beautiful
example of Penrose's idea that relativistic field equations
should reduce to holomorphicity conditions in twistor space.

The anti-self-duality condition is preserved by proper conformal
transformation, and so it makes sense to look for solutions that are
invariant under subgroups of the conformal group. The twistor
construction then gives a correspondence between {\em conformal
reductions} of the anti-self-duality condition and {\em equivariant
vector bundles} on twistor space---that is, holomorphic bundles that
are unchanged by the action of the corresponding subgroup of ${\rm
PGL}(4,\C)$. If the subgroup is $m$-dimensional and acts freely on an
open subset of $\CP_3$, then the reduced system has $4-m$ independent
variables.  The one-dimensional examples give various monopole
equations and the two-dimensional ones lead to a variety of familiar
and widely-studied intergable systems---the KdV equation, the
nonlinear Schr\"odinger equation, the Ernst equation, and many
others. The three-dimensional ones give systems of ODEs with the
Painlev\'e property.  One thus sees in the twistor geometry a very
direct connection between integrability of systems of partial
differential equations and the Painlev\'e property of the systems of
ODEs derived from them by dimensional reduction.

The Yang-Mills twistor construction encompasses only one special class
of integrable systems. But the general ideas extend to include
others---whether or not they can be taken far enough to include {\em
all}~ integrable systems is an open question, and one that is unlikely
to be answered so long as `integrability' and `twistor' retain their
current elasticity of meaning.  Whatever form the extension takes,
however, one expects to see systems of ODEs with the Painlev\'e
property at the foot of any chain of dimensional reductions.  One
reason for this is the connection between the Painlev\'e property and
isomonodromy, and the existence of a very general geometric
construction for isomonodromic families of ODEs.

\section{Equivariant bundles and isomonodromy}
Suppose that we are given
\begin{itemize}
\item a complex manifold $\cZ$ and a 
complex Lie algebra $\g$ of the same dimension
that acts on $\cZ$, with the action being free on the complement of a
hypersurface $\Sigma\subset \cZ$;
   
\item a $\g$-equivariant holomorphic principal bundle $P\to
\cZ$ with structure group $\hat G$;

\item an embedded copy $X\subset \cZ$ of $\CP_1$ which intersects
$\Sigma$ transversally, and which has the properties that $P\vert_X$ is
trivial and that 
$$
H^0(N,X)\neq 0, \qquad H^1(N,X)=0\, ,
$$
\end{itemize}
where $N$ is the normal bundle of $X$ in $\cZ$.  An action of $\g$ on
$\cZ$ is a Lie algebra homomorphism into the holomorphic vector fields
on $\cZ$; the action is free at a point $z$ if corresponding map $\g
\to T_z\cZ$ is injective, and therefore, on dimensional grounds, an
isomorphism.  Each element of $\g$ determines a holomorphic vector
field $Y$ on $\cZ$.  When $P$ is equivariant, then these vector fields
in turn lift to vector fields on $P$ that are preserved by the action
of $\hat G$.  In a local trivialization, the lift is given by a linear map 
$Y\mapsto \theta_Y$, where $\theta_Y$ is a function on $\cZ$ with
values in the Lie algebra $\hat \g$, with the property
$$
Y(\theta_{Y'})+Y'(\theta_{Y})+[\theta_Y,\theta_{Y'}]=0
$$
for every pair of generators $Y,Y'$ of the $\g$ action.  Under gauge
transformations of the local trivialiation, $\theta_Y\mapsto
h^{-1}\theta_Yh+h^{-1}Y(h)$, where $h$ takes values in $\hat G$.

By Kodaira's theorem \cite{K}, $X$ is one of family of curves $X_m\subset
\cZ$ that intersect $\Sigma$ transversally, 
labelled by a parameter space $M$ of dimension $H^0(X,N)$; for almost
every $m\in M$, the restricted bundle $P\vert_{X_m}$ is
trivial---although this does not imply that $P$ itself is trivial.  
The {\em jumping lines} are the isolated members of the family for which 
$P\vert_{X_m}$ is nontrivial.

Given these, we construct an isomonodromic family of ODEs as follows.
First, we note that the action of $\g$ on $P$ determines a flat
$P$-connection $D$ on the complement of $\Sigma$, characterized by
$D_Y=\rd +\theta_Y$ for each generator $Y$.
Next,
$P$ and the connection are pulled back to the {\em correspondence
space} $\cC$. This is the space whose points are pairs $(z,X_m)$, with
$X_m$ one of the family of curves and $z\in X_m$.  If we exclude the
jumping lines from $M$, then the pull-back of $P$ is the trivial
bundle, and the pull-back of the connection is flat and meromorphic:
it is singular where $z\in \Sigma$.

The correspondence space is fibred over $\cZ$ by $(z,X_m)\mapsto z$,
and over $M$ by $(z,X_m) \mapsto m$; the fibres of the second
fibration are the curves $(z,X_m)$, with $m$ fixed, which are all 
copies of $\CP_1$. In the global trivialization of
the pulled-back bundle over $\cC$, the restriction of the connection to
one of these fibres is of the form
$$
\rd - A(\zeta)\, \rd \zeta
$$
where $\zeta$ is a stereographic coordinate and $A$ is a rational
function on $\CP_1$ with values in $\hat\g$---the Lie algebra of $\hat
G$. It has poles at the points 
where $X_m$ meets $\Sigma$ .  Thus we have a family of linear ODEs,
labelled by points of $M$
$$
\frac{\rd y}{\rd \zeta}=Ay
$$
Here $y$ takes values in a representation space of $\hat G$. The ODEs
are uniquely determined by the data up to conjugation of $A$ by a
holomorphic map $h:M\to \hat G$. 

If the coordinate is chosen so that $\zeta=\infty$ is not an
intersection point with $\Sigma$, then $A$ has a zero of order 2 at
infinity.  A pole of order $r+1$ in $A$ is a singularity of {\em rank}
$r$ in the linear system.

The solutions to the ODE are parallel sections of the associated vector
bundle over the lines $X_m$---although these exist only locally and
are not single-valued in the large.  Isomonodromy follows more or less
directly from the fact that $D$ is the restriction of a flat
meromorphic connection.  If all the poles of $A$ are simple,
then `isomonodromic' means no more than that the monodromy
representation is constant up to conjugacy as $m$ varies; if there are
poles of higher order, then it involves in addition the preservation
of other data assocatied with the ODEs.  In either case, the monodromy
representation coincides with the holonomy of the flat connection on
$P$.

The archetypical example is the sixth Painlev\'e equation \cite{MW2}.  This we
obtain by taking $\cZ$ to be a neighbourhood of a line in $\CP_3$ and
$\g$ to be the diagonal subalgebra of the Lie algebra of the
projective gneral linear group ${\rm PGL}(4,\C)$.  The solutions to
$P_{VI}$ correspond to the equivariant
${\rm SL}(2,\C)$-bundles over $\cZ$.

\section{Local form of the connection}

In a local trivialization of $P$,
\be
D=\rd -\alpha
\label{locform}\ee
where $\alpha$ is a meromorphic 1-form on $\cZ$ with values in the
$\hat g$. It is nonsingular on the complement of 
$\Sigma$ and satisfies the flatness condition
$$
\rd \alpha-[\alpha,\alpha]=0\, .
$$
Its restriction to a curve $X_m$ is gauge-equivalent to $A\, \rd
\zeta$.

We shall make two `genericity' 
assumptions.
The first is that there is an abelian subalgebra
$\ft\subset \hat g$ such the leading coefficient in $A$ is conjugate
to an element of $\hat g$ at each pole.  When $\hat G$ is
the general linear group and $\ft$ is the diagonal subalgebra, this is
a consequence of the the standard genericity asumption that the
eigenvalues of the leading coefficients are distinct. The second assumption is
that there is a curve in the family through every point of $\cZ$.

Let $a\in \Sigma \cap X$.  By the second assumption, we can identify a
neigbourhood of $a$ in $\cZ$ with a neighbourhood of $(a,0)$ in
$\Sigma\times \C$ the curves $\zeta\mapsto (z,\zeta)$, with $z$ fixed,
are parts of curves in the family.  Suppose that $A$ has a pole of
order $r+1$ at $a$.  Then, by the first assumption and the flatness
condition,
$$
h\alpha h^{-1}=\frac{a\,\rd
\zeta +\zeta \beta}{\zeta^{r+1}} +\gamma\, ,
$$
where $h$ takes values in $\hat G$ and is holomorphic, $a$ takes
values in $\ft$, $\beta$ and $\gamma$ are holomorphic at $\zeta=0$,
$\beta$ has no $\rd
\zeta$ component, and
$$
\rd_z a=-r\beta+O(\zeta^{r+1})\, ,
$$
where $\rd_z$ denotes the exterior derivative with $\zeta$ held
fixed.  By expanding $a$ in powers of $\zeta$, we deduce that
\be
h \alpha h^{-1}=\rd \tau +m\, \rd \log\zeta
+\gamma'\, ,
\label{tau}\ee
where $\tau=p/\zeta^r$, $p$ a $\ft$-vlaued polynomial in $\zeta$, $m$
is a constant element of $\ft$, and $\gamma'$ is holomorphic at
$\zeta=0$.  The constant $m$ is the `exponent of formal monodromy',
and the restriction of $\tau+m\log \zeta$ to a one of the curves of
the family is the diagonal exponent in the formal soultion of the
linear system in \cite{JMU}.

If $Y$ is one of the  generatiors of  the action of $\g$,
then $i_Y\alpha$ is constant and $Y(\zeta)=0$ on $\Sigma$.  So, despite
the fact that $\tau$ blows up as $\zeta^{-r}$, its
derivative $Y(\tau)$ is holomorphic at $\zeta=0$.

\section{Two constructions}

In general, the information contained in the linear system of ODEs and
its deformations is contained in the action of $\g$ on both the base
space $\cZ$ and the bundle $P$. At the extremes are two special
constructions.  

\subsection*{First construction}

In the first, due to Hitchin \cite{H}, $P$ is trivial and all the data are
encoded in the geometry of $\cZ$ and $\g$.  Hitchin's construction has
been exploited to generate interesting geometries by using Penrose's
nonlinear graviton construction, and its variants.  With $\g={\rm
sl}(2,\C)$, for example, $\cZ$ is the twistor space of a
four-dimensional complex Riemannian manifold with anti-self-dual
confromal structure and $SL(2,\C)$ symmetry (a `Bianchi IX' geometry).
It is shown in \cite{Wo} that if $\g$ is the Lie algebra of the general
linear group and if $A$ satisfies the standard condition that its
leading coefficients at its irregular singlarities have distinct
eigenvalues, then the corresponding isomonodromic family can be
obtained from such a twistor space.

\subsection*{Second construction}

At the other extreme, $\cZ$ is a standard space that carries only
information about the number of singularities in the linear system and
their ranks.  The particular isomonodromic family is encoded in the
bundle $P\to \cZ$.  Such a standard twistor space can be constructed
as follows. Let $\ft$ be an abelian subalgebra of $\hat \g$---we shall
keep in mind the main
example in which $\ft$ is the Lie algebra of the diagonal subgroup of 
the complex general linear group. Let $r$ be a nonnegative 
integer. If $r=0$,  let
$H_r$ denote the abelian group $\C^{*}$ (the multiplicative group of complex
numbers ); if $r>0$, let $H_r=\C^{*}\times \ft$, with the group
law
$$
(\lambda ,t).(\lambda',t')=(\lambda \lambda',\lambda^r
t'+\lambda^{\prime r}t).
$$
For $r>0$, this acts linearly on the vector space $V_r=\C\oplus \ft$ by
$$
(Z,W)\in \C\oplus \ft \mapsto (\lambda Z, \lambda^r W+tZ^r)\, .
$$
For $r=0$, we take $V_0=\C$.

Denote by $H$ the quotient group $(H_{r_0}\times \cdots \times H_{r_n})/\C^{*}$.
The {\em standard twistor space} $\cZ_S$ associated with $H$ 
is constructed from the linear 
action of the $H_{r}$s on
$$
\bV =V_{r_0}\oplus \cdots \oplus V_{r_n}\, ,$$
where the $r_i$s are the ranks of the singularities of the linear
system.  By picking out the $Z$s in each summand, we have a projection
$\pi:\bV \to \C^{n+1}$. 
We define $\cZ_S$ to be the quotient of $$\bV\setminus \pi^{-1}(0)$$ 
by the action of $\C^{*}$, and take $G=H$.

As a complex manifold, $\cZ_S$ is the total space of copies of the line
bundles $\cO(r_i)$ over $\CP_n$, with the projection onto $\CP_n$ given by
$\pi$; in the Fuchsian case, in which all the ranks are zero, $\cZ_S=\CP_n$. 
The action of the $H_r$s on the $V_r$s gives an action of $H$ on
$\cZ_S$ which is transitive and free on the
complement of the  
$$
\Sigma=\bigcup \pi^{-1}(\Pi_i)/C^{*}\, ,
$$
where the $\Pi_i$s are the coordinate hyperplanes in $\C^{n+1}$, and the
curves $X_m$ are the sections of $\cZ_S$ over lines in $\CP_n$.  

Consider the form of $\alpha$ in a neighbourhood of a point on $\Pi_i$.
If $r_i>0$, then the generators of the $H$-action are nonzero
in the neighbourhood, except for those in the Lie algebra of
the $\ft$ factor of $H_{r_i}$.  Denote by $Z_i,W_i$ the corresponding 
homogeneous coordinates on $\cZ_S$. By choosing the local
trivialization of $P$ to be invariant along the non-vanishing
generators, we can ensure that $\tau$ in depends only on $Z_i,W_i$.
Moreove, since it is constant along the non-vanishing generator of $H_{r_i}$,
it depends on these variables only through the combinations
$W_i/Z^r_i$.  Consequently $\tau=p/Z^r_i$, where $p$ is a linear
function of $W_i$.  By constant linear transformation of the
variables $W_i$, therefore, we can ensure that $\tau=W_i/Z^r_i$.
This relates the coordinates homogeneous $W,Z$ on $\cZ_S$ to the
deformation parameters in \cite{JMU}.

It is
shown in \cite{Wo} that if $\hat G$ is the general linear group and if $A$
satisfies the standard condition that its leading coefficients at its
irregular singularities have distinct eigenvalues, then the
corresponding isomonodromic family can be obtained from an equivariant
bundle $P\to \cZ_S$, at least locally, that is, by restricting $\cZ_S$ to
an open neighbourhood of one of the curves $X_m$.  The ranks of the
singularities are the integers $r_i$.

The second twistor construction allows one to see, in a systematic
way, how the different isomonodromic deformation
equations fit into the hierarchies associated with standard integrable
systems.  In the archetypical example, $n=3$ and all the ranks are
zero, so $\cZ_S=\CP_3$.

\section{From the second construction to the first}

Suppose that we are given $P\to \cZ$ as in the first constrcution. The
passage from the second construction to the first is the `switch map'
in \cite{MW}, which interchanges the roles of $\g$ and $\hat\g$.  The total
space of the principal bundle $P\to \cZ$ carries actions of the $\hat
g$ (the Lie algebra of its structure group) and of $\g$ (since it is
equivariant).  The two actions commute and their
orbits are transverse except over $\Sigma$.  The
curves in $\cZ$ lift to $P$ by the triviality condition.  If we take a
neighbourhodd of such a lifted curve, and quotient by the action $\hat
\g$, then we obtain the twistor space of the second cosnstruction; the
projections from $P$ of the lifts of the  curves  in $\cZ$ are the
twistor curves in the new twsitor space.

\section{From the first construction to the second}

To go in the other direction, we start with $\cZ$ and the
$\g$-action of the first construction, with $P$ is trivial.  Then
$D$ is globally of the form (\ref{locform}).

The hypersurface $\Sigma$ is made up of a number of connected
components, each intersecting a twistor curve 
$X$ at one of the singularities of the
linear system.  Let us concentrate for the moment on one component $S$
on which $\zeta=0$. We shall construct an equivariant bundle $B_r\to
\cZ$ from $S$ and an action of $\g$ on $B_r$ that together encode
the position of the corresponding pole of $A$ along with information
about the singular part of $A$. First we define $L_r$ to be the line
bundle with $-S$.  This is equivariant because $S$ is
invariant.  It has fibre coordinate $z$ away from $S$, and fibre
coordinate $z'$ in a neigbourhood of $S$, with transition rule
$$z'=z/\zeta\, ,$$ 
and a $\g$-invariant section over the complement of $S$ given by
$z=1$. This section determines the action $\g$ on the whole of $B_r$ since
$Y(\zeta)/\zeta$ is nonsingular on $S$.

When $r>0$, we also introduce an equivariant affine bundle $T_r\to \cZ$ 
with fibre $\ft$ and
transition rule
$$
w=w'-\tau
$$
$w,w'\in \ft$ are fibre coordinates.  
The action of $\g$ is determined away from $S$ by the condition that
$z$ and $w$ should be invariant.  It extends holomorphically over $S$
since if $Y$ is one of the generating vector fields of the action of
$\g$ on $\cZ$, then
$$
\frac{Y(\zeta)}{\zeta}\quad \mbox{and}\quad  Y(\tau)
$$
are nonsingular at $\zeta\to 0$. 

In both cases, the action of $\g$ on $L_r$ and on $T_r$ 
commutes with the natural
action of $H_r$ given by
$$
z\mapsto \lambda z, \qquad w\mapsto w+t\, .
$$

We construct such a line bundle $L_{r_i}$ 
for each component of $\Sigma$, and affine bundle  $T\to \cZ$
by taking the product of the $T_r$s for each component of $\Sigma$ for
which $r>0$.  We  use a subscript $i$ to denote the
quantities associated with the $i$th component of $\Sigma$
Put
$$
P=T\times (L_{r_0}\oplus  \cdots \oplus L_{r_n})/\C^{*}\, ,
$$
where $\C^{*}$ acts by $z_i\mapsto \lambda z_i$.  The fibres of $P$
are products of $\CP_n$, with homogeneous coordinates $z_i$, with
$n+1$ copies of the affine space modelled on $\ft$. 
We can lift a twistor curve $X\subset \cZ$
to $P$ by choosing a stereographic coordinate $\lambda$ on $X$, and
putting $z_i:z_j=(\lambda-\lambda_i):(\lambda-\lambda_j)$, where the
points  
$\lambda =\lambda_i$ are the intersection points of $X$ with
$\Sigma$; and by picking out the unique section of $T\vert _X$ for
which $w_i-\tau_i$ is holomorphic at the intersection with $S_i$.

We then commuting actions of $\g$ and $H$ on $P$.  We also have a
projection $P\to \cZ_S$ given by
$$
(z_i,w_i)\mapsto (Z_i,W_i)=(z_i,z_i^rw_i)
$$
which extends holomorphically to the fibres of $P$ over $\Sigma$
since $z=\zeta z'$ and $z^rw=z^{\prime r}\zeta^r(w'-\tau)$ are
holomorphic at $\zeta=0$.   Moreover, if we lift a twistor curve to
$P$ and then project it into $\cZ_S$, then we obtain one of the
twistor curves in $\cZ_S$.

The quotient $P=B/\C^{*}$ by
the subgroup $\C^{*}\subset H$ is
part of the total space of a principal bundle with structure group
$H/\C^{*}$
over a neigbourhood $U$ of a twistor curves in the standard
twistor space. From it, we recover the whole of $P$ over $U$, and 
thus the second description of the isomonodromic family.

\end{document}